\documentclass[]{aa}
\usepackage{natbib}  
\usepackage{graphicx}
\usepackage[varg]{txfonts}
\usepackage{epstopdf}
\usepackage{amsmath}
\usepackage{sidecap}
\newcommand{\unitHflux}{Mx$^2$ cm$^{-2}$ s$^{-1}$}

\begin{document} 

\newcommand{\asr}{ {\it Ann. Rev. Astron. Astrophys.}}
\newcommand{\jaa}{ {\it J. Astrophys. Astron.}}
\newcommand{\sungeo}{ {\it Sun Geosph.}}
\newcommand{\jfm}{ {\it J. Fluid Mech.}}
\newcommand{\cpcf}{{\it Comments Plasma Phys. Controlled Fusion}}

   \title{Successive Injection of Opposite Magnetic Helicity in Solar Active Region NOAA 11928}

   \author{P. Vemareddy\inst{1} \and P. D\'emoulin\inst{2}
               }

   \institute{Indian Institute of Astrophysics,
              II Block, Koramangala, Bangalore-560 034, India \\
              \email{vemareddy@iiap.res.in} \and
			  Observatoire de Paris, LESIA, UMR 8109 (CNRS), 
			  F-92195 Meudon, France \\
              \email{Pascal.Demoulin@obspm.fr}
							 }
   
	\titlerunning{Successive Injection of Opposite Magnetic Helicity in AR 11928}
	\authorrunning{Vemareddy and Demoulin}
	\date{Received: July 7, 2016; Accepted: November 1, 2016 }

  \abstract
{}
{Understanding the nature and evolution of %boundary driven 
the photospheric helicity flux transfer is a key to reveal the role of magnetic helicity in coronal dynamics of solar active regions. 
}
{Using SDO/HMI photospheric vector magnetograms and the derived flow velocity field, we computed boundary driven helicity flux with a 12 minute cadence during the emergence of AR 11928. Accounting the foot point connectivity defined by non-linear force-free magnetic extrapolations, we derived and analyzed the corrected distribution of helicity flux maps.  
}
% Results
{The photospheric helicity flux injection is found to changes sign during the steady emergence of the AR. This reversal is confirmed with the evolution of the photospheric electric currents and with the coronal connectivity as observed in EUV wavelengths with SDO/AIA.  During about the three first days of emergence the AR coronal helicity is positive while later on the field configuration is close to a potential field.  As theoretically expected, the magnetic helicity cancelation is associated to enhanced coronal activity.
}
{The study suggests a boundary driven transformation of the chirality in the global AR magnetic structure. This may be the result of the emergence of a flux rope with positive twist around its apex while it has negative twist in its legs. The origin of such mixed helicity flux rope in the convective zone is challenging for models. 
}

\maketitle
%%----------------------------------------------------------------------------------------
\section{Introduction}

The magnetic energy and helicity in the solar active regions (ARs) are two important parameters for a quantitative study of magnetic origins of solar eruptions. The field lines in a closed magnetic structure of the corona have foot points rooted in the photospheric boundary. During AR emergence and their evolution, the lower boundary acts as a driver of the evolution in the structure either via boundary flows or via the injection of additional structure through it. 

Magnetic helicity is a metric describing the volumetric complexity like twist and shear magnetic field of the AR.
It is a well conserved quantity even in non-ideal cases and its injection, under ideal conditions, through boundary-like photosphere, $S$, is derived \citep{berger1984, finn1985} as
\begin{equation}
{{\left. \frac{dH}{dt} \right|}_{S}} = 2\int\limits_{S}{\left( {{\mathbf{A}}_{P}}\bullet {{\mathbf{B}}_{t}} \right){{\text{v}}_{n}}dS}-2\int\limits_{S}{\left( {{\mathbf{A}}_{p}}\bullet {{\mathbf{v}}_{t}} \right){{\text{B}}_{n}}dS}   \,,
\label{eq1}
\end{equation}
where $\mathbf{A}_p$ is the vector potential of the potential field $\mathbf{B}_p$ computed from the photospheric $B_n$ distribution, $\mathbf{B}_t$ and $B_n$ are the tangential and normal magnetic fields, and $\mathbf{v}_t$ and $v_n$ are the tangential and normal components of the plasma velocity $\mathbf{v}$.  This relation identifies that the magnetic helicity in the corona primarily originates from the twisted magnetic flux tubes emerging from the solar interior into the corona (first term; $v_n$ term hereafter), and is further generated by shearing and braiding the field lines by the tangential motions on the solar surface (second term; $v_t$ term). While flux is emerging, using a geometrical relation of the apparent horizontal foot point velocity of field lines (${\bf u}$; flux transport velocity) with the plasma velocity ($\mathbf{v}$) implies the relation 
\begin{equation}
\mathbf{u} = \mathbf{v}_t - \frac{v_n}{B_n} \mathbf{B}_t   \,.
\label{eq2}
\end{equation}
With this equation \citet{demoulin2003} combined the two terms in Eq.~(\ref{eq1}) to measure helicity flux using the observations of line-of-sight magnetic fields at the photosphere with only the knowledge of $B_n$ and $\mathbf{u}$. 

Equation~(\ref{eq1}) with the help of~(\ref{eq2}) can be written as \citep{berger1988, pariat2005};                            
\begin{equation}
{{\left. \frac{dH}{dt} \right|}_{S}}=\frac{-1}{2\pi }\int\limits_{S}{\int\limits_{{{S}'}}{\frac{d\theta (\mathbf{x}-\mathbf{{x}'})}{dt}}}{B_{n}}(\mathbf{x}){B_{n}}(\mathbf{{x}'})dSd{S}'
\label{eq3}
\end{equation}
where
\begin{equation}
\frac{d\theta (\mathbf{x}-\mathbf{{x}'})}{dt}=\frac{{{\left. \left[ \left( \mathbf{x}-\mathbf{{x}'} \right)\times \left( \mathbf{u}-\mathbf{{u}'} \right) \right] \right|}_{n}}}{{{\left| \mathbf{x}-\mathbf{{x}'} \right|}^{2}}}
\label{eq4}
\end{equation}
is the relative angular velocity between field line foot points located at $x$ and $x'$. Here the helicity flux $dH/dt$ is the summation over all the photospheric elementary flux pairs of their net angular rotation around each other weighted by $B_n.B_n'$. For example, if two positive (negative) end points rotate counter-clock wise ($d\theta /dt>0$) then their net contribution to dH/dt is negative and consequently the field lines above becomes twisted in a left-handed sense. 

Employing a technique for deriving flow velocity, the \textit{differential affine velocity estimator for vector magnetograms} \citep[DAVE4VM;][]{schuck2008},  \citet{liuy2012} recently found that Equations (\ref{eq1}) and (\ref{eq3}) are not yielding identical results when applied to photospheric observations. Then, they suggested to calculate individual terms in Eq.~(\ref{eq1}). With that, on re-expansion of Eq.~(\ref{eq3}) with Eq.~(\ref{eq2}) yields

\begin{align}
   \left. \frac{\text{dH}}{\text{dt}} \right| _{S}
 =& \frac{1}{2\pi } \int\limits_{S} \int\limits_{S'}  \mathbf{\hat{n}} \bullet 
        \frac{\mathbf{x}-\mathbf{{x}'}} {\left| \mathbf{x}-\mathbf{{x}'} \right|^2}  
        \times \, \mathbf{bv}_n  ~dS\,d{S}'   \nonumber \\ 
 -& \frac{1}{2\pi }\int\limits_{S}\int\limits_{S'}  \mathbf{\hat{n}} \bullet 
       \frac{\mathbf{x}-\mathbf{{x}'}} {\left| \mathbf{x}-\mathbf{{x}'} \right|^2}  
        \times \, \mathbf{bv}_t  ~dS\,d{S}'   \nonumber \\ 
 =& \int\limits_{S} \left[ 
         G_{\theta ,~v_n} (\mathbf{x}) + G_{\theta ,~v_t} (\mathbf{x})  
                   \right] \, dS \nonumber \\ 
 =& \int\limits_{S} G_{\theta } (\mathbf{x}) ~dS 
\label{eq5}
\end{align}
with 
\begin{eqnarray} 
\mathbf{bv}_n &=& 
   \mathbf{B}_{t}(\mathbf{x})    ~\text{v}_{n}(\mathbf{x}) 
                                 ~\text{B}_{n}(\mathbf{{x}'})
 - \mathbf{B}_{t}(\mathbf{{x}'}) ~\text{v}_{n}(\mathbf{{x}'}) 
                                 ~\text{B}_{n}(\mathbf{x}) \nonumber \\
\mathbf{bv}_t &=& 
    \left[ \mathbf{v}_t (\mathbf{x}) - \mathbf{v}_{t} (\mathbf{{x}'}) \right]
   ~\text{B}_n (\mathbf{x}) ~\text{B}_n (\mathbf{{x}'})  \nonumber 
\end{eqnarray}
where $G_{\theta ,~v_n}$ and $G_{\theta ,~v_t}$ denote helicity flux distribution due to $v_n$ and $v_t$ terms as in Eq.~(\ref{eq1}). The above equation includes helicity injection due to apparent relative rotation of elementary polarity with regard to surrounding polarities and due also to inherent twist while flux emerges from sub-photosphere.  

\begin{figure*}[!ht]   
\centering
\includegraphics[width=.96\textwidth]{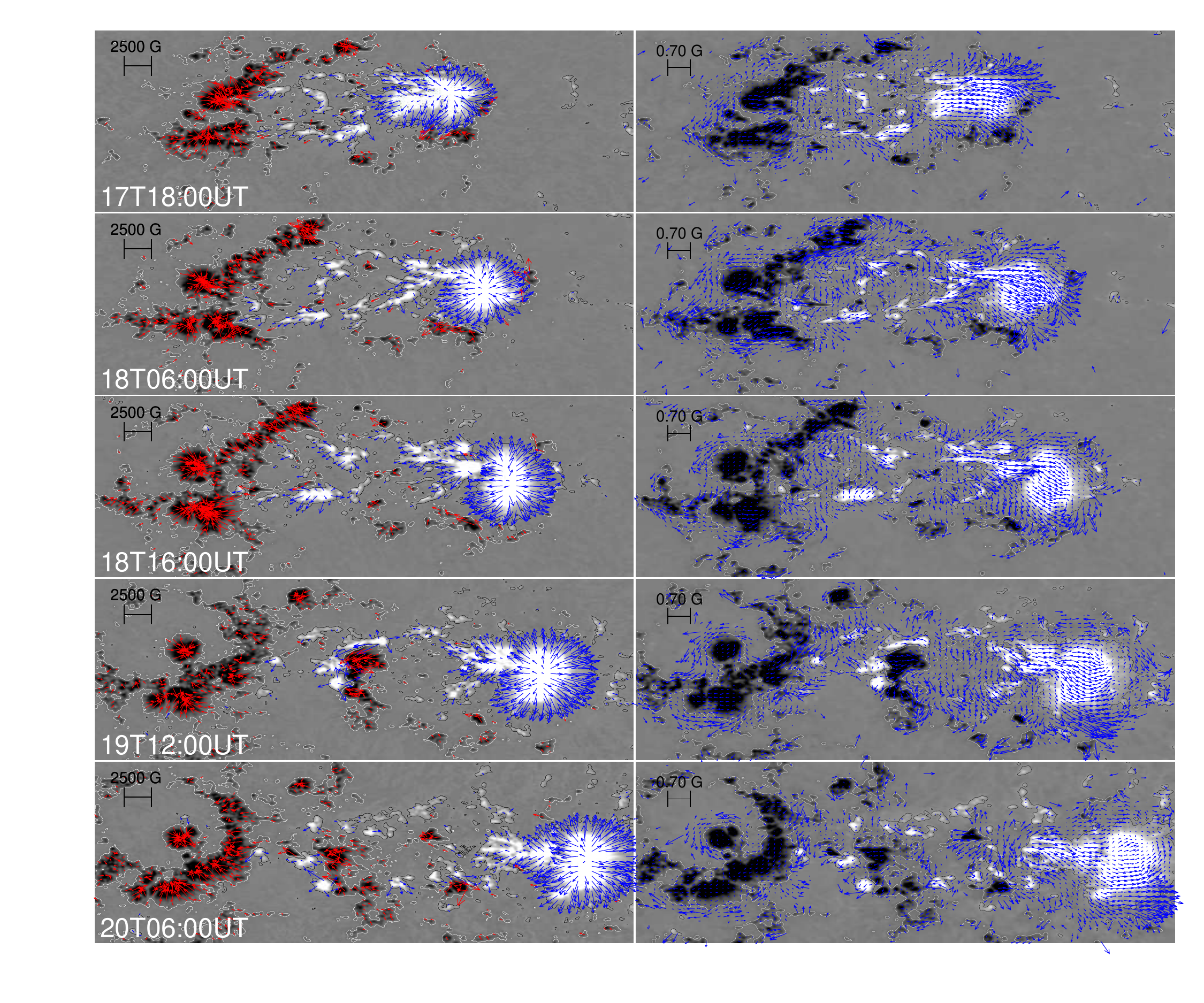}
\caption{ HMI magnetic and deduced velocity fields of AR 11928 at different times.  In all the panels, the background of grey levels shows the vertical component, $B_{z}$, of the magnetic field and contours of $B_z\pm 120$ G and the magnitude of the transverse vectors is indicated by the light blue segment located in the top left of each panel.  
   {\bf Left panels:} Vector magnetic field observations (blue/red arrows above positive/negative $B_z$ values). Vectors point to horizontal field direction and their length corresponds to the horizontal field magnitude
  (a zoom on a computer screen is needed to better view these arrows).
   {\bf Right panels:} The horizontal velocity vectors (${\bf V}_h$) are shown with blue arrows. 
   %displayed on vertical component of field. 
The velocity pattern in the leading polarity indicates the presence of a counter-clock rotating motion of the magnetic elements on 19 and 20 December 2013.  The field-of-view is a portion of actual observation indicated with white rectangular box in Fig.~\ref{Fig2}a.
%\pc{Why removing the labels showing the X,Y sizes? Would be better to have them as in Fig. 6, especially for a first figure.  They need to be large enough and the units can be added here in the caption.}\pv{V: this is a cutout from Fig.5 indicated with rectangular box in top left panel. Now the axes units are changed to arcsec in Figs. 5 and 8, but when plotting vector arrows in that coordinate system, there encounter a problem with normalisation which I am at present not dwell much. For further reasons of axes labels, I just suppress coordinate system, which is in pixels. This I am referring to Fig. 5}
}
\label{Fig1}
\end{figure*}

\begin{figure*}[!htp]
\centering
\includegraphics[width=.9\textwidth]{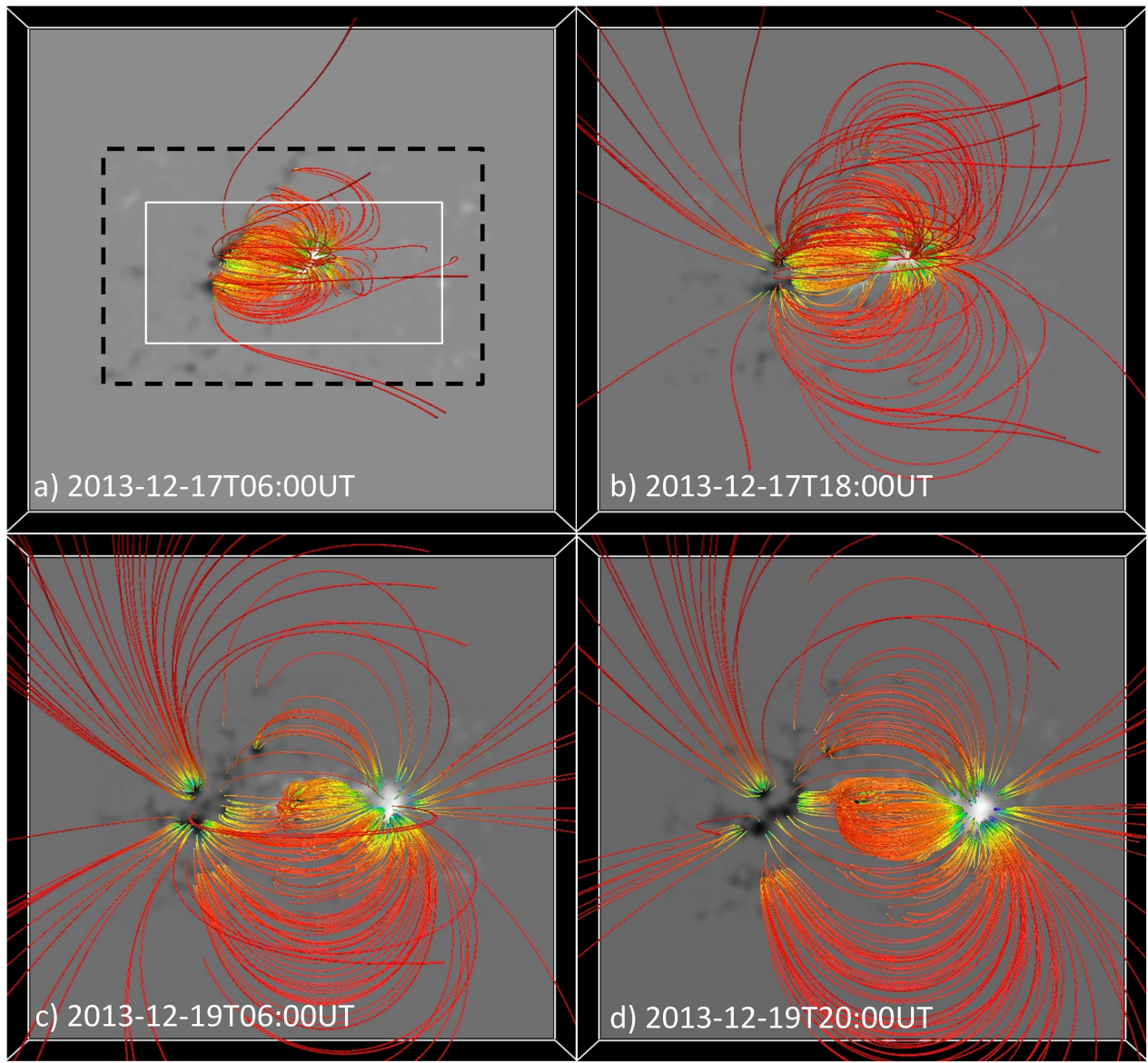}
\caption{Top view of the magnetic structure of AR 11928 at four different times of the evolution. Color scale along each field line represents the horizontal field strength. The lower boundary  (z=0) shows the vertical component $B_z$ at the photospheric level with grey levels in the range [-100,100]~G.   The white rectangular box in panel (a) refers to the field-of-view of panels in Figure~\ref{Fig1} and the black dashed rectangular box indicates the field-of-view of observations as plotted in Fig.~\ref{Fig5}
}
\label{Fig2}
\end{figure*}

Observational studies on quantitative estimates of dH/dt from Eq.~(\ref{eq3}) , or equivalent formula, were carried out to reveal the role of helicity in the eruptive nature of ARs \citep{chae2001a, demoulin2002, kusano2002, moon2002, chae2004, labonte2007}. Few studies claim that the monotonous accumulation of helicity in the corona comes from magnetic domains with uniform signed distribution of helicity flux and therefore are prone to launch coronal mass ejections \citep{pariat2006,vemareddy2012b,vemareddy2012a}. On the other hand, domains with opposite signed distribution of helicity flux were speculated to be related to more energetic events as they can liberate more free energy since the system can relax to a more potential field \citep{chandra2010, romano2011a, vemareddy2012a}. 

Since the magnetic helicity is not a local quantity, the helicity flux distribution is only meaningful when one considers a whole elementary flux tube rather than its individual foot points \citep{pariat2005}. As a consequence, the earlier interpretations of observed activity based on computed helicity flux distribution remains speculative. 
   
\begin{figure*}
\centering
\includegraphics[width=.98\textwidth]{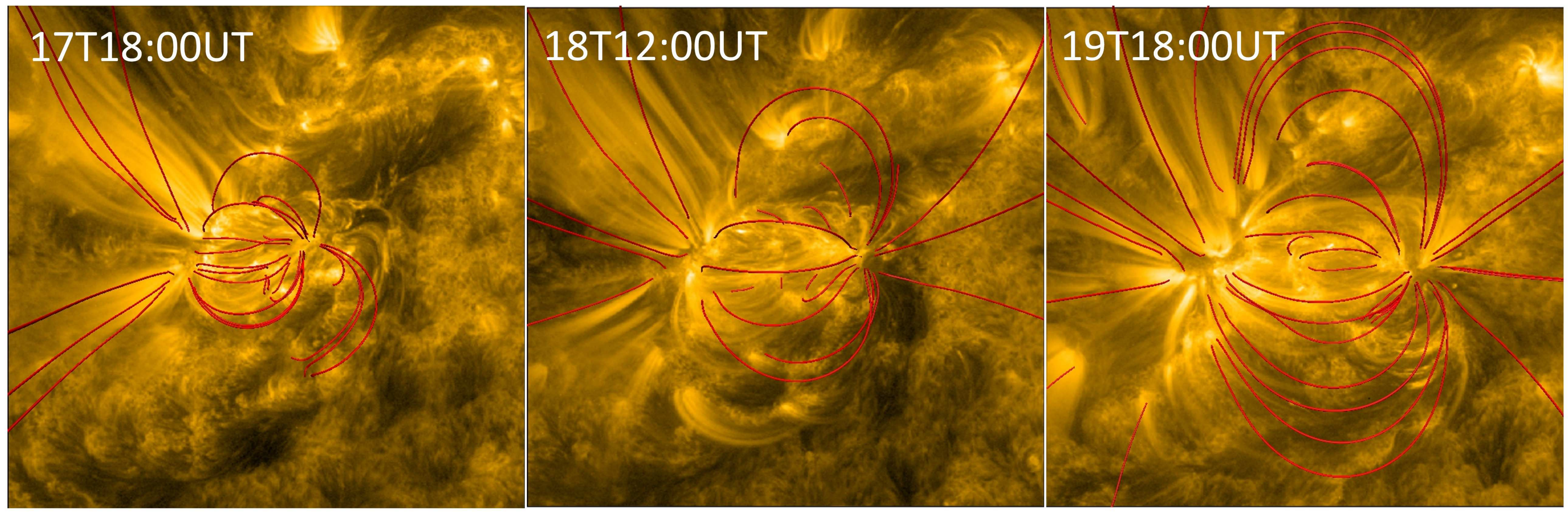}
\caption{ Field lines, computed from the magnetic extrapolation, plotted on co-aligned AIA 171~\AA\ images. 
In the core of the AR, there is a relatively good global correspondence between field lines and loops.  
% Large scale field line structure resembles the coronal loops, however deviation be noted due to missing currents in the observations at small scale. 
The field-of-view is same as Fig.~\ref{Fig2}.}
\label{171_nlff}
\end{figure*}

Considering the whole elementary flux tube, \citet{pariat2005} defined an improved helicity flux distribution assuming information of coronal connectivity (inferred from a model). The new flux density of helicity writes
\begin{equation}
G_{\Phi }({{\mathbf{x}}_{c\pm }})=\frac{1}{2}\left( G_{\theta }\left( {{\mathbf{x}}_{c\pm }} \right)+\left| \frac{{B_{n}}\left( {{\mathbf{x}}_{c\pm }} \right)}{{B_{n}}\left( {{\mathbf{x}}_{c\mp }} \right)} \right|G_{\theta }\left( {{\mathbf{x}}_{c\mp }} \right) \right)
\label{eq_gphi}
\end{equation}
where c denotes a closed elementary flux tube with foot points at $\mathbf{x}_{c\pm}$ in the photosphere. 
The factor $1/2$ present in the above equation assumes that the helicity is injected equally between the two footpoints of each field line \citep[see][ for a more general case]{pariat2005}. Recently, this helicity flux proxy was tested using analytical case studies and numerical simulations by \citet{dalmasse2014} and they observationally found evidence for the existence of opposite signed helicity flux distribution in the flaring AR 11158 \citep{dalmasse2013}. 

Present study is an attempt towards further understanding the distribution of helicity injection flux in an emerging AR. In an extensive study of three emerging ARs \citet{vemareddy2015a} found corresponding signatures of helicity flux distribution $G_{\theta }$ with the observed activity. We consider here one of their cases, AR 11928, and computed the connectivity-based helicity flux distribution ($G_{\Phi }$) at successive stages of evolution to explore its nature and a possible physical significance. Data sets and methodology are given in Sect.~\ref{Obs}, results are presented in Sect.~\ref{Res}.  A discussion of the results is made in Sect.~\ref{Disc}. 

\section{Observations and Employed Procedure}
\label{Obs}

The studied AR 11928 appeared on 16 December 2013 at a location of E40$^o$S15$^o$ on the solar disk. It emerged progressively and evolved to a leading major positive polarity sunspot and following plage group regions of negative polarity. We covered this AR evolution with 12-minute cadence magnetograms for four days since its emergence.

The required photospheric vector magnetic field observations ($\mathbf{B}$, at a resolution of 0.5\arcsec per pixel) are obtained from \textit{Helioseismic and Magnetic Imager} (HMI; \citealt{schou2012}) on board Solar Dynamic Observatory. HMI science team had pipe-lined the process of retrieving vector field information from filtergrams (\citealt{hoeksema2014}, and references therein) and made the direct usable vector products  (\texttt{hmi.sharp.cea.720s}) available to solar community. The pipeline procedure involves inversion of stokes vectors using the very fast inversion of the Stokes vector algorithm \citep{borrero2011, centeno2014} based on the Milne-Eddington atmospheric model and  removing $180^o$ azimuthal ambiguity using minimum energy method \citep{metcalf1995,leka2009}. The projection effects in the field components in the cutout area are corrected by transforming to disk center using the cylindrical equal area projection method \citep{calabretta2002, bobra2014}. Detailed documentation on the pipeline processing of HMI magnetic field data, including various data artifacts, is available in \citet{hoeksema2014}. 

From these vector magnetic field observations, we first derived the flow velocity ($\mathbf{v}$) by employing DAVE4VM technique, then the flux transport velocity (${\bf u}$) with DAVE technique. Then, we calculated $G_{\theta }(\mathbf{x})$ (\citealt{pariat2006,liuy2012, vemareddy2015a}). For connectivity information of foot points, we performed nonlinear force-free field (NLFFF) extrapolation by an optimization procedure involving minimization of Lorentz force and divergence conditions (\citealt{wiegelmann2004, wiegelmann2010}, and also see \citealt{vemareddy2014}).
The photospheric boundary conditions are derived from the vector magnetic field observations after pre-processing in order to satisfy force-free conditions the best possible \citep{wiegelmann2006}.
We embedded the lower boundary field in a large null array to minimize effect of lateral boundaries. We then re-binned the data by a factor two to have a reasonable computation time. Having done these systematic procedures, we obtained 3D magnetic field in an uniform spacing Cartesian grid of $400\times400\times256$ corresponding to physical dimensions of $292\times292\times187$~Mm$^3$ encompassing the AR. Given well known difficulties to construct 3D fields from boundary field observations (e.g., \citealt{derosa2009}), we regard these NLFFF extrapolations as an approximation to the coronal magnetic fields of this AR.

We adopted the procedure detailed in \citet{dalmasse2013} to compute $G_{\Phi }$ from Eq.~(\ref{eq_gphi}). First we traced field lines having one foot point in a polarity region ($\mathbf{x}_{c+}$/$\mathbf{x}_{c-}$) and found the corresponding conjugate foot point ($\mathbf{x}_{c-}$/$\mathbf{x}_{c+}$) after landing to the boundary at $z=0$. Having foot point coordinates of all field lines in a magnetic structure of AR, we implement Eq.~(\ref{eq_gphi}) by using a bi-linear interpolation. Traced field lines that touch the lateral boundaries are considered as open-like. In all our magnetogram sequence, the fraction of open-like field lines found is up to 10\% and the redistribution of helicity density, Eq.~(\ref{eq_gphi}), is not applied as connectivity is undefined for them. 

\section{Results}
\label{Res}
\subsection{Global Evolution}
\label{Global}

% general description of the AR evolution
AR 11928 initially emerges with a bipolar field distribution and evolves to a large leading positive polarity sunspot and a more dispersed following negative polarity. In Fig.~\ref{Fig1}, we plot the vector magnetograms at different times in the local solar frame (local horizontal and vertical directions) after the transformation from the observed frame. For both velocity and magnetic fields, the horizontal field component is plotted with arrows indicating the direction and magnitude. The background is the map of the vertical magnetic field  component ($B_z$). These panels show the emergence of small bipoles which are the consequence of the development of the undulatory instability or of the upward convective motions, creating a sea-serpent configuration with magnetic dips \citep[e.g.][]{pariat2004,valori2012}. The opposite polarities diverge and the like polarities coalesce to form strong concentrated spots, of opposite magnetic polarity which separate as the AR evolves in time. This evolution is typical for AR emergence \citep{vandriel2015}. While these prominent polarities are in separating motion, a reverse orientation bipole, larger than others, forms and develops in the middle of the AR (panel at 19/12:00 UT in Fig.~\ref{Fig1}). This creates a complex magnetic topology for the AR coronal configuration (Fig.~\ref{Fig2}c). This bipole progressively disappeared with time (e.g., the bottom panel at 20/06:00 UT in Fig.~\ref{Fig1}).

In order to compare the NLFF model for the coronal field, we plot the field lines on AIA 171~\AA\ observations in Figure~\ref{171_nlff}. Field lines in the core of the AR globally resemble the coronal loops.
However, a closer look shows also some deviations (e.g. the computed field lines are more symmetric in the East-West direction than the observed loops as some show a sharp bend near the leading polarity). Such deviations could have several origins. This could be due to the missing electric currents in the magnetograms due to a too coarse spatial resolution not able to resolve their magnetic structures. 
This effect is further increased by the rebinning procedure applied to the magnetograms in order to achieve feasible computing times \citep[e.g.][]{derosa2015}. Moreover, convergence to a NLFF field with a small divergence for the magnetic field is another issue \citep[e.g.][]{wiegelmann2012}. Finally, the involved assumptions with the boundary observations are not yet settled \citep[e.g., the magnetograms are not in a fully force-free region in contrast to the equations solved in the coronal volume,][]{derosa2009}.  All these issues contribute to the deviations of the model field lines with the actual observations. Despite all these limitations we obtained a coherent global resemblance between modeled field lines and the coronal loops over the entire AR evolution.       

% Derive V
In order to examine the magnetic helicity flux, we derived the photospheric velocity field by employing DAVE4VM method on these time sequence of vector magnetic field observations. The horizontal velocity field at different stages is plotted on the $B_z$ map in the right panels of Fig.~\ref{Fig1}. The velocities are typically up to 0.9~km/s and earlier on they are dominantly translational in the leading polarity.  However, by December 18 onwards, the velocity vectors in the leading positive polarity indicate the presence of an anticlockwise rotating motion especially in the southern portion of the leading polarity. This motion becomes prominent in later time while separation motion continues (see the three right bottom panels of Fig.~\ref{Fig1}).       

\begin{figure}
%\sidecaption
\includegraphics[width=.48\textwidth]{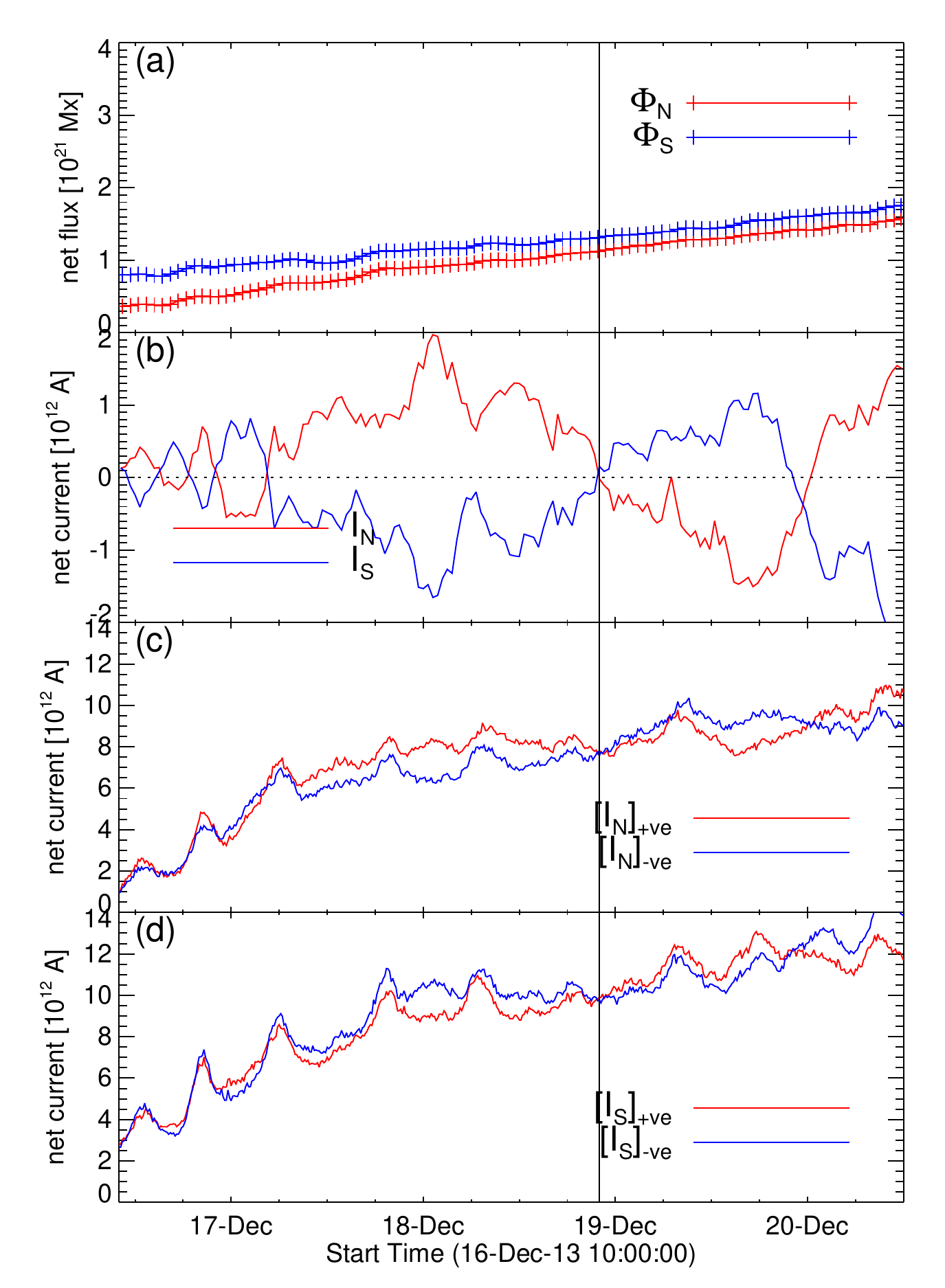}
\caption{Evolution of magnetic flux and electric current in AR 11928. 
  (a) Net magnetic flux ($\Phi$) from north (N, $B_z>0$) and south (S, $B_z<0$) polarity as a function of time. 
  (b) Total net vertical current (I) and net currents from north (I$_{\rm N}$) and south (I$_{\rm S}$) magnetic polarity as a function of time.  The vertical line at 18T22:00 UT marks the sign reversal of net current in each magnetic polarity.
  Panels (c) and (d) separate absolute value of positive and negative currents ($+ve$ and $-ve$) from north and south polarities, respectively. 
  }
\label{Fig3}
\end{figure}

% Extrapolation
In Fig.~\ref{Fig2}, we plot the extrapolated magnetic structure of AR 11928 at different stages of evolution. The field lines are represented with a color scale of horizontal field strength. The magnetic structure is non-potential in the early emergence phase and becomes more relaxed as AR evolves. December 18 onwards, the field structure appears close to potential field, globally connecting the leading and following polarities. Due to the formation of a reverse orientation bipolar region in between the separating leading and following polarity, a low lying bipolar structure exists in the December 19 frames. 

% Magnetic flux
We derived the net vertical magnetic flux $\Phi =\sum\limits_{i=1,N}{{{\left( {B_{z}} \right)}_{i}}\,\Delta x\Delta y}$ ) in both AR magnetic polarities (Fig.~\ref{Fig3}a). The flux of both polarities predominantly increases in time owing to the emergence of these flux regions. This AR has an almost constant rate of flux emergence which is not so frequent (e.g., see examples in \citealt{poisson2015}). The observed activity is limited to GOES class C without coronal mass ejections (CMEs). However, jet like ejections often occurred during the AR evolution.  

 \subsection{Electric Current Evolution}

% Global current evolution
We also computed net vertical electric current ($I=\sum\limits_{i=1,N}{{{\left( {{J}_{z}} \right)}_{i}}\,\Delta x\Delta y}$) in each AR magnetic polarity and plot their time evolution in Fig.~\ref{Fig3}b. Here $J_{z}$ in the local frame is derived with the horizontal field components
\begin{equation*}
J_z=\frac{1}{\mu_0}\left(\frac{\partial B_y}{\partial x}-\frac{\partial B_x}{\partial y}\right)
\end{equation*}
where $\mu_0=4\pi\times10^{-7}H\,m^{-1}$. Partial derivatives are approximated by a three-point Lagrangian interpolation procedure. The net current in each polarity is much more time varying both in magnitude and sign than the magnetic flux (Fig.~\ref{Fig3}a,b). 
The current profiles show prominent variations with a time scale of around 12 hours, while they are small in the magnetic flux profiles. They are due to orbital rotation of the SDO spacecraft, and are also seen in other active region studies \citep{hoeksema2014}.  There is also a longer time scale variation at the scale of a day which is comparable in the north and south polarities (as expected if the field is force-free). The absolute value of both currents first reach maximum at 02:00 UT on December 18 ($2\times10^{12}$A in the north polarity and $-1.8\times10^{12}$A in south one). Later on the net currents exhibit a counter evolution, then they both change in sign at 22:00 UT on December 18 and with again a reversal in sign around the end of December 19. 

% Almost neutralisation
Since the positive/negative magnetic polarity is predominantly due to leader/follower polarity, we can also interpret the current evolution in terms of unbalanced current in the leader/following polarity. For that, we separately compute the positive ($I_+$) and negative currents ($I_-$) in each polarity and compared their absolute values in Fig.~\ref{Fig3}c,d.  Those currents are about a factor 10 larger in magnitude than those described before (Fig.~\ref{Fig3}b).
The ratio of positive and negative current varies within 0.82-1.33 in the leader polarity, and within 0.83-1.18 in the follower polarity, then there is nearly neutralization of currents in each magnetic polarities. This is due to a small magnetic shear along the main AR inversion line (Fig.~\ref{Fig1}, left panels) as shown by \citet{dalmasse2015}.

\begin{figure}[!ht]
\centering
\includegraphics[width=.49\textwidth]{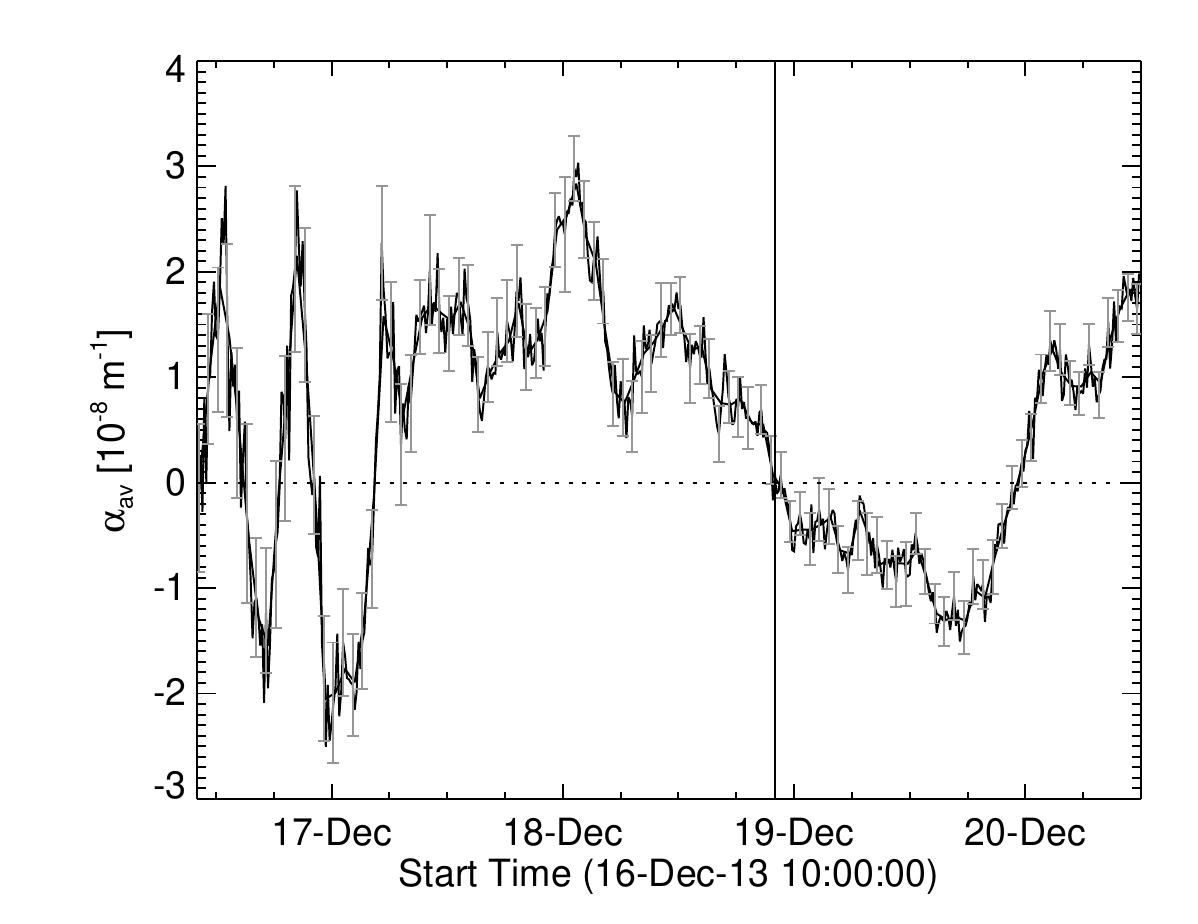}
\caption{Time evolution of $\alpha_{av}$ (defined by Eq.~\ref{eq7}) in the entire AR 11928. The error bars are obtained from a least square regression procedure in the plot of $J_z$ and $B_z$ (see Eq.~\ref{eq8}). $\alpha_{av}$ changes sign several times at the beginning of the emergence, as well as later on when the AR is well developed at 18T22:00 UT (vertical line). } 
\label{Fig4}
\end{figure}

% alpha evolution
We next investigate the mean evolution of $\alpha = J_{z}/B_{z}$ called $\alpha_{av}$. It is a proxy for the average twist of the AR field lines \citep{hagino2004}. Using the three components of vector magnetic field, we compute this parameter as  
\begin{equation}
	{\alpha }_{av}=\frac{\sum{{J_{z}}(x,y)\text{sign}[{B_{z}}(x,y)]}}{\sum{|{B_{z}}|}}
\label{eq7}
\end{equation}
If the positive and negative  magnetic fluxes are in balance, the above equation is equivalent to 
\begin{equation*}
\alpha_{av}=\frac{1}{2}\left[\frac{\sum\limits_{B_z>0}J_z(x,y)}{\sum\limits_{B_z<0}B_z(x,y)}+\frac{\sum\limits_{B_z>0}J_z(x,y)}{\sum\limits_{B_z<0}B_z(x,y)}\right]
\end{equation*} 
which is the mean of $\frac{\sum J_z}{\sum B_z}$ over the two polarities. It is an estimation of how twisted is the magnetic configuration.
Since Eq.~(\ref{eq7}) corresponds to the usual least-squares fit assuming a linear regression $J_z=\alpha_{av}B_z$, the error in $\alpha_{av}$ is estimated by
\begin{equation}
 \delta \alpha _{av}^{2}=\frac{{{\sum{\left[ {{J}_{z}}(x,y)-{{\alpha }_{av}}{B_{z}}(x,y) \right]}}^{2}}/|{B_{z}}(x,y)|}{(N-1)\sum{|{B_{z}}(x,y)}|}
\label{eq8}
\end{equation}	
where N is the number of pixels with $|{\bf B}_t|>150$~G. 
	
The time variation of ${\alpha }_{av}$ is shown in Fig.~\ref{Fig4} with the error bars indicating $\pm\delta\alpha_{av}$. Just like the net current in the early emergence phase (Fig.~\ref{Fig3}b), $\alpha_{av}$ changes sign four times. This is likely due to the limited spatial resolution since the AR has a low spatial extension at that times. After 04:00 UT on December 17, it remains positive with significant variations till 22:00 UT on December 18 (vertical line), from which time it turns negative, reversing of sign again by the end of December 19. $\alpha_{av}$ is indeed related to the current distribution in the AR but with a different weighting in the summation than for the net currents. 
%SCfigure is used
\begin{figure*} 
\sidecaption
\includegraphics[width=.6\textwidth]{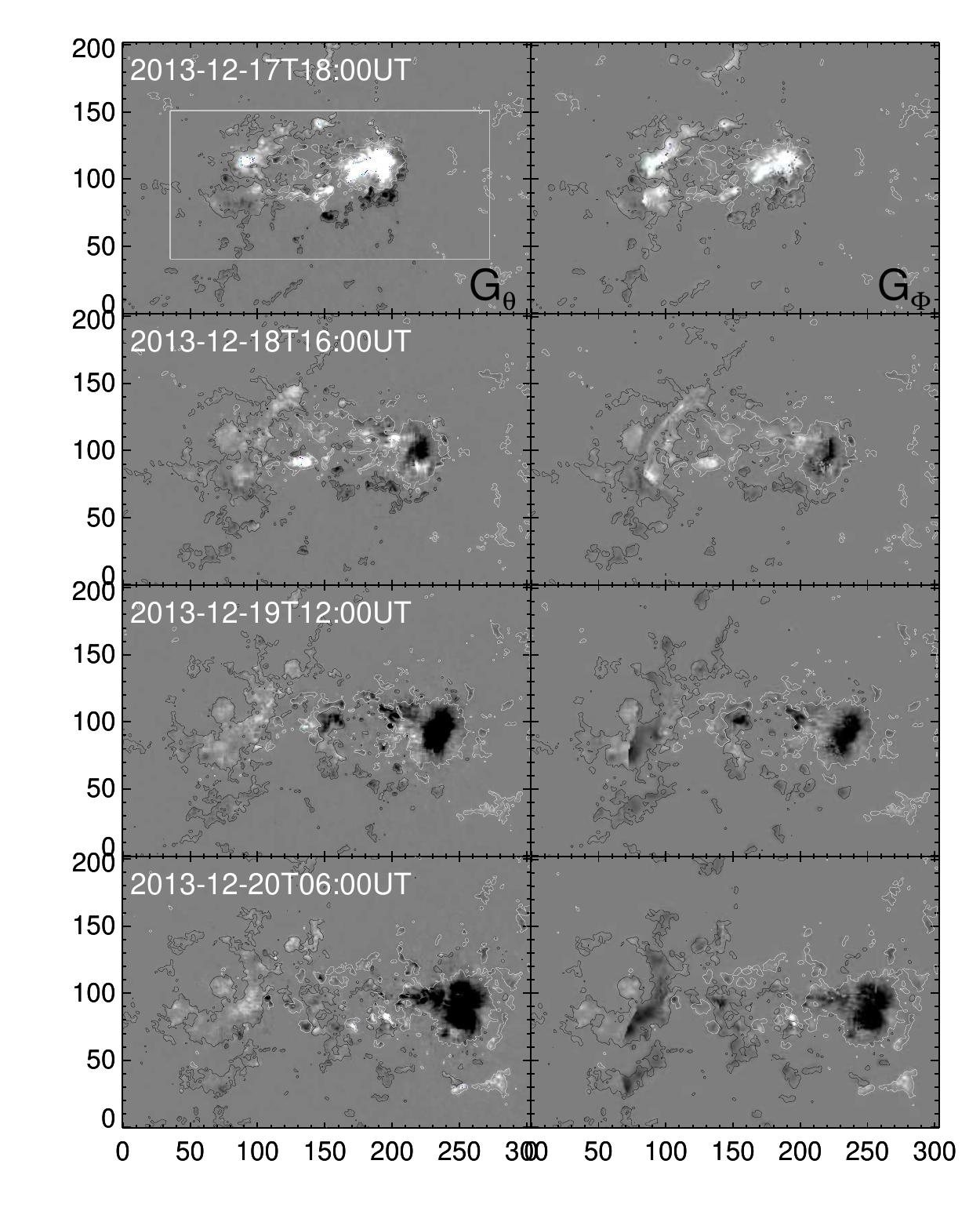}
\caption{Snapshots of helicity flux distributions with $G_\theta$ maps (left column) and $G_\Phi$ maps (right column) at four different times of the evolution of AR 11928. $G_\Phi$ is derived from $G_\theta$ taking into account the connectivity of the field line foot points (Eq.~(\ref{eq_gphi})). Intense positive helicity flux in $G_\theta$ maps is associated with the leading major sunspot polarity, which turns to negative over the time. This flux is re-distributed to the following polarity flux by closed field lines in $G_\Phi$ maps. In all panels, contours of $B_z$ at $\pm120$~G (black/white) are over plotted. All maps were scaled with grey levels within $\pm 1\times10^{-19}$ \unitHflux . 
Axes units are in arc-second and the field-of-view is same as indicated by the black dashed rectangle in Fig.~\ref{Fig2}a. The rectangular white box in the top left panel refers to the field-of-view of the panels in Figure~\ref{Fig1}. }
\label{Fig5}
\end{figure*}

\subsection{Helicity Flux}
% description of $G_{\theta }$
The flux density of magnetic helicity is computed with the method recalled in Sect.~1. Figure~\ref{Fig5} shows the helicity flux distribution at four times of Fig.~\ref{Fig1}. $G_{\theta }$ distribution (left column panels) shows both positive and negative values. The distribution has a range of flux exceeding $\pm 1\times10^{19}$ \unitHflux , but we scaled the maps within these values.  A dominant positive signal is persistent mostly within the leading positive sunspot till 06:00 UT on December 18, by which time the sunspot got well separated from the following polarity. In the later stages, strong negative signal with the leading sunspot kept increasing in magnitude. 
 
% magnetic tongues
On December 17 and 18, the magnetic polarities are elongated and are globally resembling a yin yang pattern.
These observed features, called magnetic tongues, are produced by the azimuthal field component of the emerging flux rope projected on the vertical direction (see e.g. \citealt{poisson2015}). They are present when the magnetic flux is growing as long as the top horizontal portion of the twisted flux tube is crossing the photosphere. The magnetic-flux distribution due to the magnetic tongues is directly related to the sign of the twist in the emerging AR (see Fig.~1 in \citealt{luoni2011}).  In AR 11928 magnetic tongues on December 17 and 18 indicates a positive twist in agreement with  the positive helicity flux. In contrast, the negative helicity flux observed next has no trace on the distribution of the vertical field component.

% Description of $G_{\Phi }$
Having defined connectivity from NLFFF extrapolation (Sect.~2, Fig.~\ref{Fig2}), the computed $G_{\Phi }$ distribution is shown in the right column of Fig.~\ref{Fig5}. It is computed with Eq.~(\ref{eq_gphi}) when the field line is closed within the computation box, otherwise the local value $G_{\theta }$ is kept. Since field lines from the leader sunspot connect to the following polarity, the dominant helicity flux from the leading sunspot redistributes to the following polarity regions by means of Eq.~(\ref{eq_gphi}). Since the helicity flux is larger in absolute value in the leading polarity than the following ones, the $G_{\Phi }$ distribution is also negative in the following polarity on 19 December and later on. 

\begin{figure}[!htp]
\centering
\includegraphics[width=.48\textwidth]{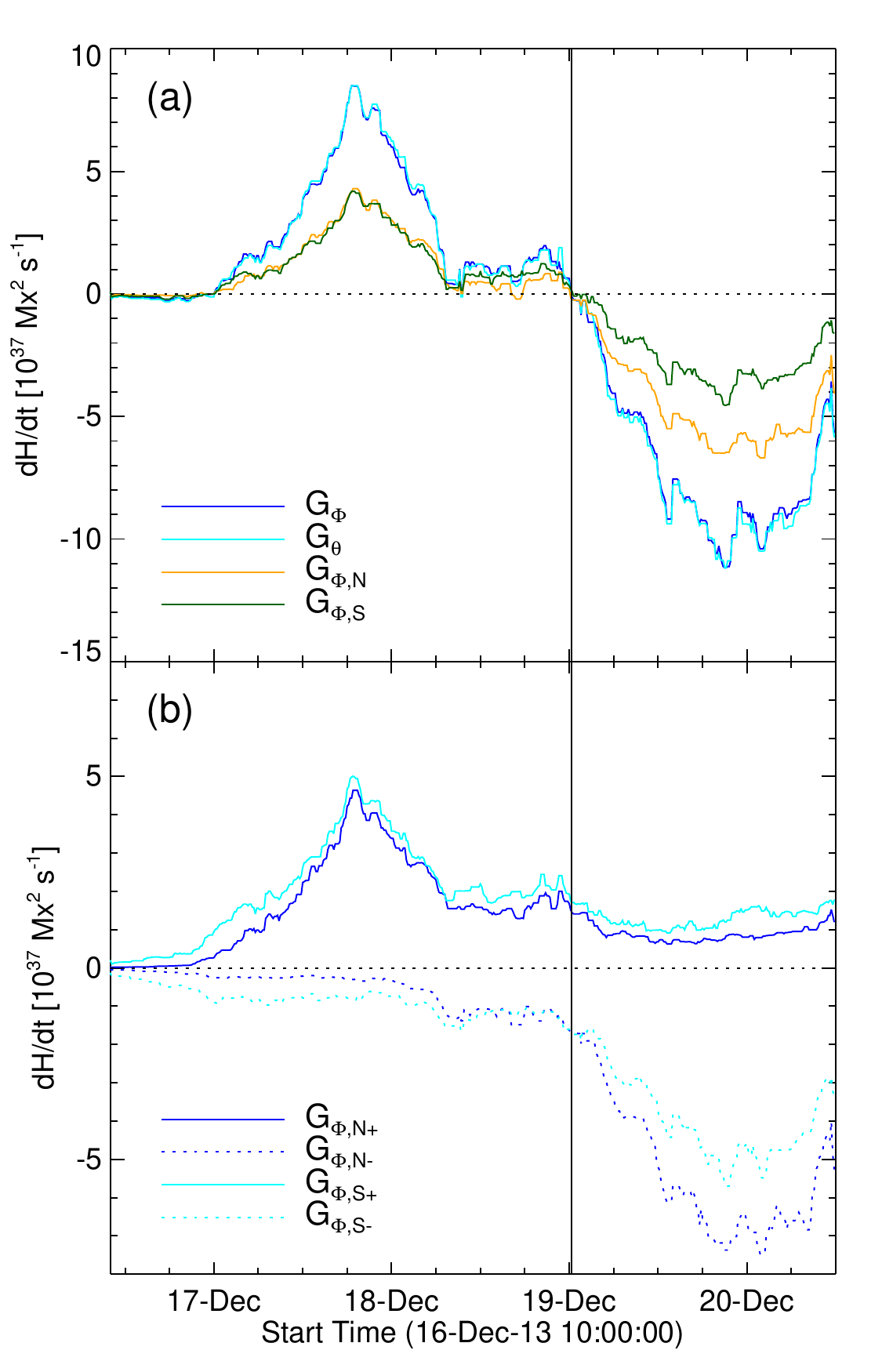}
\caption{{\bf (a)} Time evolution of the net helicity flux in AR 11928, which turns from positive to negative at 20T00:20 UT (black vertical line).  
Earlier on the net magnetic helicity flux computed with $G_{\Phi }$ have well correlated profiles in both north (orange) and south (green) polarities. This indicates a proper redistribution at foot points of closed field lines (reaching the computation box boundary).  Later on, after 18:00 UT December 18, unequal net values in north and south polarities are present. It is due to undefined connectivity for open field lines. 
{\bf (b)} Evolution of the signed net helicity flux in positive and negative magnetic polarities.  
}
\label{Fig6}
\end{figure}

%Time profiles
In Fig.~\ref{Fig6}a, we plot the summation of $G_{\theta }$ and $G_{\Phi }$ over the entire AR with respect to time. Both $G_{\theta }$ and $G_{\Phi }$ profiles have undergone a smoothing window of 7 successive data points. The time profiles of $G_{\theta }$ and $G_{\Phi }$ fluxes are correlated to a high accuracy (Fig.~\ref{Fig6}a) as expected from the redistribution which theoretically preserves the total helicity flux \citep{pariat2005}. The net helicity flux increases from zero to a maximum value of $8.5\times 10^{-37}$ \unitHflux\ at 18:00 UT on December 17. Owing to the change in sign of $G_{\theta }$ and $G_{\Phi }$ (Fig.~\ref{Fig5}), the helicity flux then decreases toward near zero values. Next, it turns to negative value at 00:20 UT on December 20. This helicity time profile has similar variations than the net current profiles in both polarities (Fig.~\ref{Fig3}b, taking into account that the current sign is reverse in the following negative polarity) and to $\alpha_{\rm av}$ (Fig.~\ref{Fig4}) except different relative amplitudes and a time delay (between 2.5 and 7~h depending if the reversal or the extremums are taken into account). 

% contributions of vt and vn terms
The net $G_{\theta }$ flux is dominated by the $v_t$ term (Eq.~(\ref{eq5})) over $v_n$ by a factor of about six, so the helicity flux evolution is mostly related to horizontal boundary motions \citep[see Fig.~6 of][]{vemareddy2015a}. Still, we do observed a similar flux profiles for the $v_t$ and $v_n$ terms. Moreover, while this separation in $v_n$ and $v_t$ terms was studied in many previous publications, its physical relevance is doubtful  since each term is not separately gauge invariant in Eq.~(\ref{eq1}) and a particular gauge was used to derive Eq.~(\ref{eq5}) \citep{pariat2015}.

\begin{figure}[!htp]
\centering
\includegraphics[width=.48\textwidth]{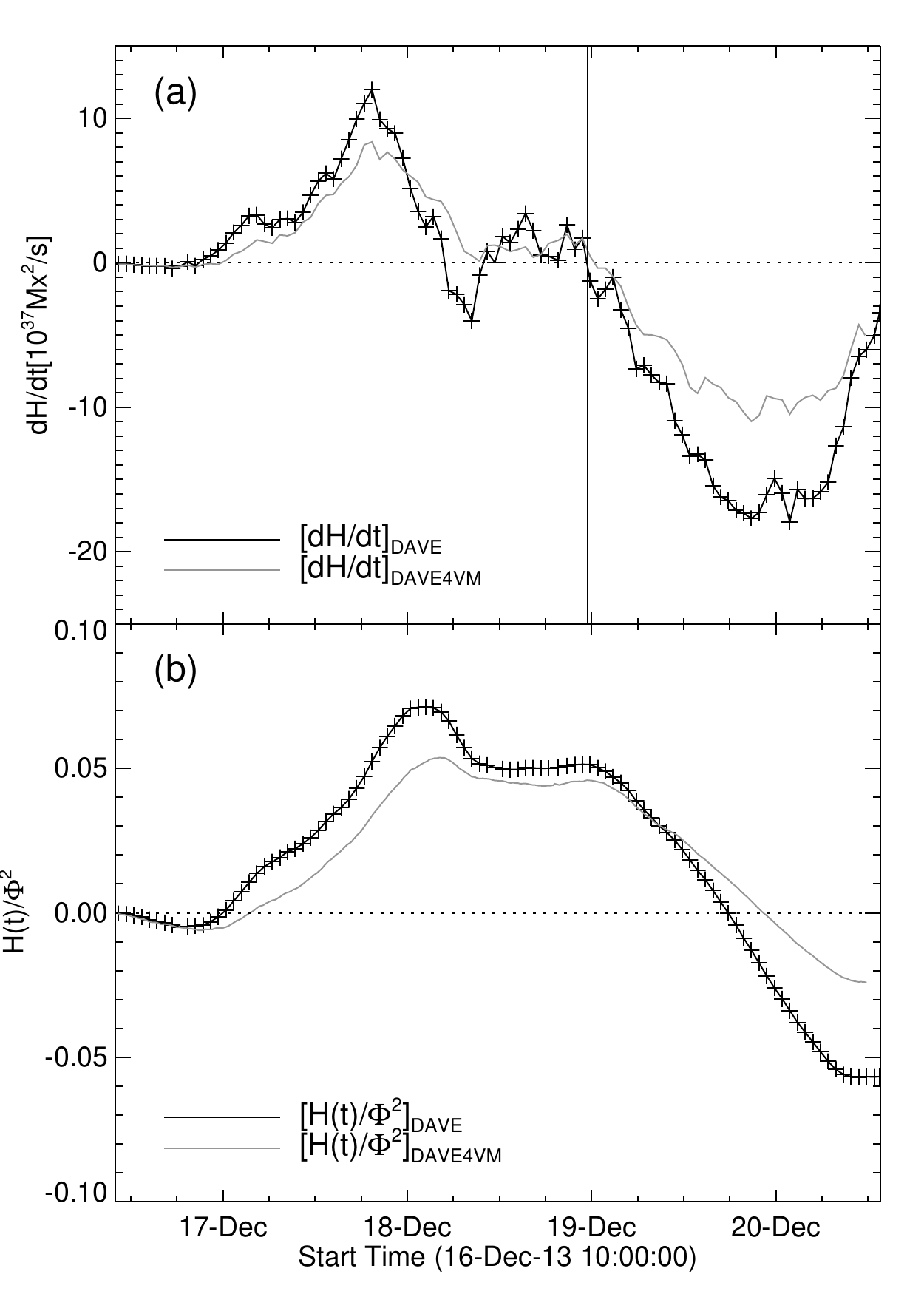}
\caption{(a) Comparison of helicity flux calculated from DAVE4VM (estimating $\mathbf{v}$, Eq.~(\ref{eq1})) and DAVE (estimating $\mathbf{u}$, Eq.~(\ref{eq3})). The general behavior and the sign reversal at 19T22:00 UT found with DAVE is similar to the one found with DAVE4VM. (b) Comparison of the accumulated helicity in the AR coronal part since mid 16 Dec. as computed with Eq.~(\ref{eq9}) and normalized by $\Phi^2$ which has the same units than helicity ($\Phi$ being the mean of positive and negative magnetic fluxes). 
 }
\label{Fig7}
\end{figure}

% describe results with $G_{\Phi }$
Furthermore, we separately computed net helicity flux of $G_{\Phi }$ from north ($G_{\Phi,~N}$) and south ($G_{\Phi,~S}$) polarity regions.  
The underlying assumption here is that the net $G_{\Phi,~N}$ and $G_{\Phi,~S}$ fluxes should be equal provided exact connectivity for every polarity pixel, i.e., closed flux system in the AR. From yellow and green curves of Fig.~\ref{Fig6}a, their time profiles have approximately equal trend till 18:00 UT on December 18, 2013. After that they exhibit noticeable deviation, with dominant helicity flux from north (leading) polarity. With increasingly separation, the open field lines (those reaching top/lateral boundaries) from leader polarity increases, then a full  redistribution is not applicable and therefore dominant negative helicity flux from N-polarity after 19 December. 
 
Next, we separately computed the net $G_{\Phi }$ flux from north, $G_{\Phi,~N}$, and south, $G_{\Phi,~S}$, polarity regions separating the positive and negative contributions (Fig.~\ref{Fig6}b). This shows that the helicity flux is a mix of positive and negative contributions with two main phases: a positive injection followed, about 2 days later, by a negative injection from both polarities. 

%comparison of dH/dt derived from DAVE and DAVE4VM
We also compute $G_\theta$ from the horizontal velocity field, $\mathbf{u}$, derived from DAVE \citep{schuck2005} using line-of-sight magnetograms.
We obtained LOS magnetic field observations from HMI and we align them to the time of the central meridian passage by removing the differential rotation.  The helicity flux is computed with Eqs.~(\ref{eq3},\ref{eq4}). Its time profile $(dH/dt)_{\rm DAVE}$ is shown in Fig.~\ref{Fig7} in a comparison to $(dH/dt)_{\rm DAVE4VM}$.  
Both profiles are well correlated in time.   
However, they differ significantly in magnitude at most times. Despite that, the sign reversal time of net helicity flux in both methods agree within 2.3 hours. 
(For DAVE4VM the reversal is at 19T00:20 UT while for DAVE it is at 18T22:00 UT).

The accumulated helicity in the corona over time is given by
\begin{equation}
H(t)=\int_{0}^{t} (dH/dt)\,dt   \,.
\label{eq9}
\end{equation}

\begin{figure*}[!htp]
\centering
\includegraphics[width=.98\textwidth]{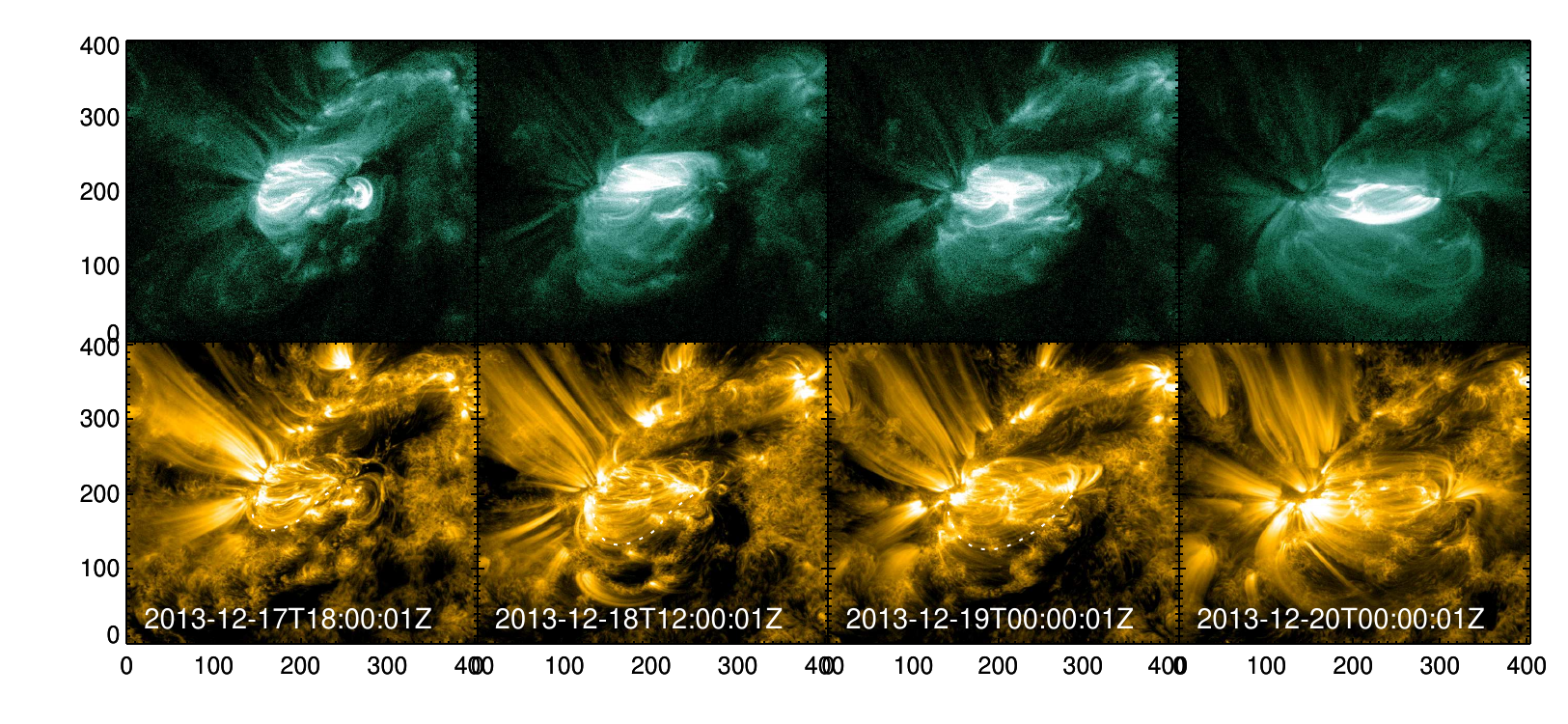}
\caption{ AIA coronal observations of AR 11928 in 94~\AA~(top row) and 171~\AA~(bottom row) wavelengths. Plasma loops represent the magnetic structure in the AR connecting the magnetic regions at the photosphere. A prominent loop and its shape transformation is shown with a dotted white line. The field-of-view is same as Figure~\ref{Fig2} with axis units in arc-seconds and origin set at the left bottom corner.} 
\label{Fig8}
\end{figure*}

Since the magnitude of magnetic helicity is proportional to square of the magnetic flux \citep{berger1984}, to compare magnetic helicity with other ARs, we plotted in Fig.~\ref{Fig7}b the normalized helicity ($H(t)/\Phi^2$) where $\Phi$ is the average AR flux between polarities ($\Phi=(|\Phi_S|+|\Phi_N|)/2$).  It indicates how much the magnetic configuration is twisted/sheared, because for a uniformly twisted flux tube with n turns, the helicity H is equal to $n~\Phi^2$, where $\Phi$ is its axial flux. In our case, the absolute value of the normalized helicity is less than 0.07 turns before and after $H$ reversal. These values are in comparison with previous AR studies having a minimum of 0.01 to a maximum of 0.2 (see for a review \citealt{demoulin2009}). The sign reversal of this quantity, for both DAVE and DAVE4VM procedures, is well after the reversal of dH/dt  (19T18:00UT) because it requires time to cancel pre-accumulated positive helicity by pumping negative quantity.  Since the positive and negative $H$ injections are realized in almost the same magnetic structures (Fig.~\ref{Fig5}), the helicity cancelation can occur, as the negative injection is occurring, without the need of magnetic reconnection (between independent flux tubes).

\section{Discussion}
\label{Disc}
Solar magnetic fields, upon their emergence, are driven by photospheric plasma motions governed by the induction equation. Consequently the coronal magnetic helicity, which plays prime role in most of the activity, is generated by these photospheric motions. In the present work, we studied the time evolution of connectivity based helicity flux of an emerging AR associated only to weak flaring activity. 

Reconstruction of coronal magnetic field every 12 minutes using observed field is a computationally expensive task. Moreover, several issues of the NLFFF modeling of the coronal field with observed photospheric boundary data are still not fully solved (see Sect. \ref{Global}).  
The implications of these issues for helicity studies can be checked by comparing the helicity flux maps and the helicity fluxes derived with and without involving the computed coronal connectivities.  Finally, the derived results are useful for studying helicity flux maps in relation to coronal activity as the studied AR shows.

The helicity flux derived from the photospheric velocities derived from DAVE4VM changes from positive to negative sign around the end of December 18. Helicity flux derived from DAVE method also exhibits a similar evolution profile. The evolution of average twist (${\alpha }_{av}$) has a comparable evolution but delayed in time by few hours (up to 7), a delay needed to accumulate enough coronal magnetic helicity. At the photosphere, plasma motions drive magnetic fields, so the observed evolution of $\alpha_{av}$ is most likely caused by these boundary evolution. Recent reports \citep[e.g.][ and references therein]{vemareddy2012b} also delineate such a relation of sunspot rotation with the nature of helicity of magnetic fields.

\begin{figure}[!htp]
\centering
\includegraphics[width=0.49\textwidth]{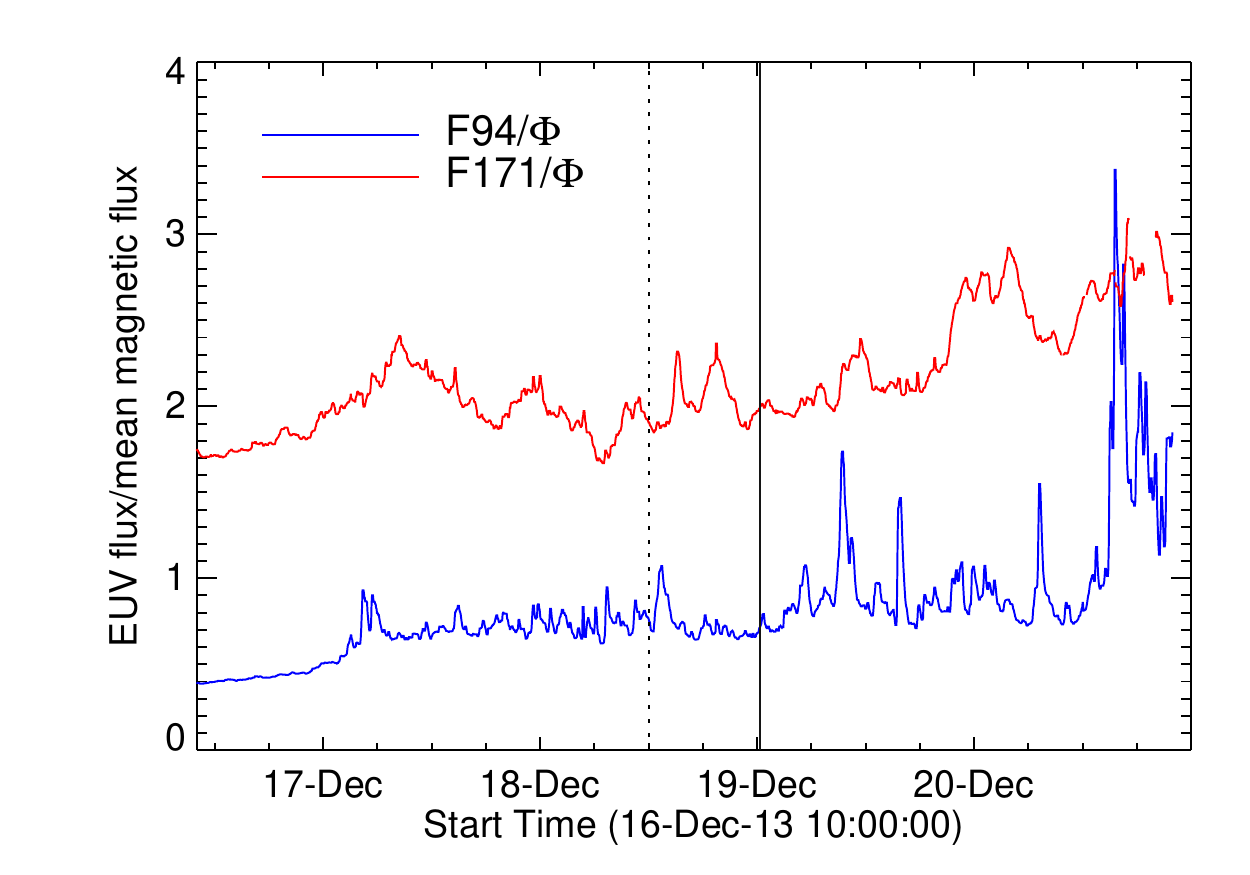}
\caption{ Time evolution of EUV flux of AIA 171 and 94~\AA~passbands. These fluxes are normalized by the mean magnetic flux ($\Phi$) of photospheric polarities to suppress enhanced trend due to emerging magnetic flux and by a constant scaling factor (arbitary units). Vertical dotted line marks the time when the AR is positioned at the central meridian and the solid vertical line refers to the sign reversal time of the photospheric helicity flux (Fig.~\ref{Fig6}a).
}
\label{Fig9}
\end{figure}

The coronal consequences of the change of helicity flux sign is present in the observed coronal loops which are tracing part of the magnetic structure. We examined the AIA coronal observations obtained in 94 and 171~\AA\ wavelengths. In Fig.~\ref{Fig8}, we plot them at four different time instances. 
During the evolution till end of 18 December, the coronal loops exhibit a S-shape morphology (e.g. the dotted line in Fig.~\ref{Fig8} frames).  This is also well present when comparing the observed coronal loops to a potential field extrapolation represented with the same viewing point. This indicates a positive magnetic helicity  in agreement with the mild clock-wise motions detected in leading polarity and the positively oriented magnetic tongues observed with the vertical field component (Fig.~\ref{Fig1}). Next, from 19 December onwards, the magnetic elements within the leading sunspot present counter-clock motions. As a response, the coronal structure becomes more potential like.  
 
The profile of normalized helicity $H/\Phi^2$ also is in the line of these coronal observations. Its maximum value is 0.07 turns so it is equivalent in term of magnetic helicity to a weakly twisted flux tube.  The injection of opposite helicity in the same magnetic structures in the $G_{\Phi }$ maps is an indication of cancellation of coronal helicity by the negative injection later on. In fact the coronal loops are not far from potential field at 19T18:00UT. 

A further clue comes from the observed activity. Rather than helicity injection of opposite sign, if helicity injection would have continue to be positive its storage in the corona would have build an increasingly stressed magnetic field which is a good candidate to become unstable and launch a CME. 
However, the observed activity is limited to C-class flares and jets so that the energy is released in many small events, and as well as coronal heating, rather than building a flux rope which is expected to erupt as a CME at some point of the evolution. 

In order to follow the coronal flaring activity we computed EUV 94 and 171~\AA\ fluxes integrated within the AR with 3 minute cadence.  These fluxes (F94, F171) are normalized by mean magnetic line-of-sight flux to suppress the increased EUV flux due to magnetic flux becoming larger \citep[e.g. see][ and references therein]{demoulin2004}.  Next, we scaled the fluxes by an appropriate constant value to plot them on the same graph because the evolution trend is the only required observable for the following analysis.
Time profiles of these fluxes are plotted in Fig.~\ref{Fig9}. Enhanced emissions in both F94 and F171 are present after the sign reversal of the magnetic helicity flux (indicated by a vertical solid black line).  The short time peaks of F94 are due to C-class flares (C1.8 at 19/09:03, C1.6 at 19/10:14, C2.2 at 19/15:26, C1.4 at 19/19:39, C5.4 at 20/16:23, C8.5 at 20/15:26, C3.2 at 20/14:52, C2.7 at 20/21:08, and C2.3 at 20/17:11UT) because AIA 94~\AA\ passband detects hot emission ($\approx~6$ MK) near soft X-ray range (1-100~\AA ). The F171 enhancement is more continuous, without peaks, indicating less hot emission over the entire AR (Fig.~\ref{Fig8}).  The coronal helicity cancellation is realized progressively by the continuous injection of opposite helicity flux (Fig.~\ref{Fig7}a). Therefore the enhancement of F94 and F171 fluxes is continuing well after mid of 19 Dec.  

In summary, the evolution of AR 11928 is peculiar compare to most of ARs studied previously since the positive H injection is followed by the larger negative injection in the same emerging magnetic structure. The initial positive injection is supported by DAVE, DAVE4VM, $\alpha_{av}$, the magnetic tongues and the shape of coronal loops. The negative injection is supported by DAVE, DAVE4VM, $\alpha_{av}$ and the shape of coronal loops which become more potential. This AR provides an example of magnetic energy due to cancelation of magnetic helicities of opposite sign. Of course in present AR the maximum value of $H/\Phi^2$ is modest so the amount of available magnetic energy is also modest. 

Finally, these results suggest a scenario of an emerging flux tube with helicity distribution changing sign over its length. This could have been created in the convective zone by a vortex locally rotating the magnetic flux tube which, by conservation of $H$, creates both $H>0$ and $H<0$ on the sides of the rotated region. In order to answer whether it is an isolated case or whether it reveals a relatively common convective zone process, the study of a much broader sample of emerging ARs is needed.

%%--------------------------------------------------
\begin{acknowledgements}
The data used here are courtesy of the NASA/SDO and the HMI science team. This work used the DAVE and DAVE4VM codes, written and developed by P. W. Schuck at the Naval Research Laboratory. We thank T. Wiegelmann for providing NLFFF code. P. V. is supported by an INSPIRE grant under the AORC scheme of the Department of Science and Technology. This work acknowledges an extensive usage of the multi-node, multi-processor high performance computing facility at IIA. We thank an anonymous referee for critical comments and suggestions.
\end{acknowledgements}
%%-------------------------------------------------
%\bibliographystyle{aa}
%\bibliography{ref_hel}

%%-------------------------------------------------
\end{document}